\documentclass[11pt,a4paper]{article}
\usepackage{jcappub,natbib}
\usepackage{braket}
\title{Observed parity-odd CMB temperature bispectrum}
\author[a,b]{Maresuke Shiraishi,}
\author[a,b]{Michele Liguori}
\author[c]{and James R. Fergusson}

\affiliation[a]{Dipartimento di Fisica e Astronomia ``G. Galilei'', \\ 
Universit\`a degli Studi di Padova, via Marzolo 8, I-35131, Padova, Italy}
\affiliation[b]{INFN, Sezione di Padova, \\ 
via Marzolo 8, I-35131, Padova, Italy}
\affiliation[c]{Centre for Theoretical Cosmology, \\ 
Department of Applied Mathematics and Theoretical Physics, \\ 
University of Cambridge, Wilberforce Road, Cambridge CB3 0WA, United Kingdom}

\abstract{%
Parity-odd non-Gaussianities create a variety of temperature bispectra in the cosmic microwave background (CMB), defined in the domain: $\ell_1 + \ell_2 + \ell_3 = {\rm odd}$. These models are yet unconstrained in the literature, that so far focused exclusively on the more common parity-even scenarios. In this work, we provide the first experimental constraints on parity-odd bispectrum signals in WMAP 9-year temperature data, using a separable modal parity-odd estimator. Comparing theoretical bispectrum templates to the observed bispectrum, we place constraints on the so-called nonlineality parameters of parity-odd tensor non-Gaussianities predicted by several Early Universe models. Our technique also generates a model-independent, smoothed reconstruction of the bispectrum of the data for parity-odd configurations. 
}

\begin{document}

\maketitle
\flushbottom

\section{Introduction}

Due to recent experimental progress in Cosmic Microwave Background (CMB) observations, it has now become possible to investigate primordial non-Gaussianities (NGs) with nearly cosmic-variance-limited accuracy \cite{Bennett:2012zja, Ade:2013ydc}. The most stringent NG constraints to date have been obtained from temperature bispectrum estimation of {\it Planck} data \cite{Ade:2013ydc}. Forthcoming analyses of {\it Planck} polarization data are expected to further improve on current limits \cite{Babich:2004yc, Yadav:2007rk}. 

All previous bispectrum estimations are based on the assumption of parity symmetry, namely $\ell_1 + \ell_2 + \ell_3 = {\rm even}$. On the other hand, some Early Universe scenarios predict parity-odd graviton NGs \cite{Lue:1998mq, Maldacena:2011nz, Soda:2011am, Shiraishi:2011st, Shiraishi:2012sn, Zhu:2013fja, Cook:2013xea, Shiraishi:2013kxa}. In such case, the resulting temperature bispectra appear in the parity-odd domain ($\ell_1 + \ell_2 + \ell_3 = {\rm odd}$) \cite{Kamionkowski:2010rb, Shiraishi:2011st, Shiraishi:2012sn,Shiraishi:2013kxa}. These theoretical predictions motivate us to study these yet unconstrained parity-odd signals in observed CMB data.

In a previous paper \cite{Shiraishi:2014roa}, we have developed a general methodology for estimating parity-odd CMB bispectra. This essentially consists in an extended version of the so-called separable modal methodology already adopted for parity-even bispectra \cite{Fergusson:2009nv, Fergusson:2010dm, Liguori:2010hx, Fergusson:2011sa, Fergusson:2014gea}. In this approach, the bispectra under study are decomposed as a sum of separable modal basis templates, and this decomposition is then exploited to achieve fast estimation by means of a KSW approach \cite{Komatsu:2003iq, Komatsu:2003fd,Creminelli:2005hu,Yadav:2007rk,Yadav:2007ny,Yadav:2007yy,Komatsu:2008hk,Senatore:2009gt,Smith:2006ud}. 

The main goal of this paper is to constrain the parity-odd bispectrum from observed temperature data by use of the parity-odd separable modal estimator \cite{Shiraishi:2014roa}. Our dataset will consist in the coadded V+W WMAP 9-year data \cite{Bennett:2012zja}. 

As we will see, the parity-odd bispectra under examination are negligible at high-$\ell$ ($\ell \simeq 250$), so that the resolution 
of the WMAP dataset already allows to get close to optimal temperature constraints (our technique is currently implemented only for temperature bispectra. The inclusion of polarized bispectra will be discussed in a forthcoming publication \cite{Liguori:2014}).

Before delving into actual data analysis, we validate our estimator by analyzing simulated NG maps with given nonlinearity parameter $f_{\rm NL}$. After this preliminary step, we move on to compute the ``modal coefficients" $\beta_n$ (see section \ref{sec:reconstruction}) and reconstruct the parity-odd temperature bispectrum from WMAP data. 

We then fit the observed bispectrum to different theoretical parity-odd shapes to obtain WMAP constraints on parity-odd $f_{\rm NL}$. We focus on three specific Early Universe models,  associated respectively with Weyl gravity \cite{Shiraishi:2011st}, a rolling pseudoscalar \cite{Shiraishi:2013kxa}, and large-scale helical primordial magnetic fields (PMFs) \cite{Shiraishi:2012sn}. The former two models predict equilateral-type graviton NGs, while the tensor NG created in the helical PMF model is amplified in the squeezed limit. 

This paper is organized as follows. In the next section, we review the parity-odd separable modal estimator, following the 
treatment of ref.~\cite{Shiraishi:2014roa}, and check its validity on simulations. In section~\ref{sec:reconstruction}, we estimate $\beta_n$ from the WMAP data, and reconstruct the WMAP temperature bispectrum. Section~\ref{sec:constraints} presents constraints on parity-odd $f_{\rm NL}$ associated with the Weyl, pseudoscalar and helical PMF models. In the final section 
we provide our comments and conclusions.

\section{Parity-odd separable modal estimator}\label{sec:modal}

In this section, we start by summarizing our implementation of an optimal modal NG estimator for parity-odd CMB bispectra~\cite{Shiraishi:2014roa}. We then check its validity using simulated NG maps, in view of the following bispectrum analysis of WMAP data.

\subsection{Analytic expression} 

An estimator for measuring the strength of primordial NG basically correlates theoretical bispectrum templates to the observed three-point function:
\begin{eqnarray}
{\cal E} &=& \frac{1}{N^2}
\sum_{\ell_i m_i}
\left(
\begin{array}{ccc}
\ell_1 & \ell_2 & \ell_3 \\
m_1 & m_2 & m_3
\end{array}
\right)
B_{\ell_1 \ell_2 \ell_3}(-1)^{\ell_1 + \ell_2 + \ell_3} 
\left[ 
\frac{a_{\ell_1 m_1} a_{\ell_2 m_2} a_{\ell_3 m_3}}{C_{\ell_1} C_{\ell_2} C_{\ell_3}}
\right. \nonumber \\   
&&\left.\quad
- \frac{a_{\ell_1 m_1}}{C_{\ell_1}} 
\frac{\Braket{a_{\ell_2 m_2} a_{\ell_3 m_3}}_{\rm MC} }{C_{\ell_2} C_{\ell_3}}
- \frac{a_{\ell_2 m_2}}{C_{\ell_2}} 
\frac{\Braket{a_{\ell_3 m_3} a_{\ell_1 m_1}}_{\rm MC} }{C_{\ell_3} C_{\ell_1}}
- \frac{a_{\ell_3 m_3}}{C_{\ell_3}} 
\frac{\Braket{a_{\ell_1 m_1} a_{\ell_2 m_2}}_{\rm MC} }{C_{\ell_1} C_{\ell_2}} \right],
\label{eq:estimator}
\end{eqnarray}
where $B_{\ell_1 \ell_2 \ell_3}$ is a given theoretical bispectrum with $f_{\rm NL} = 1$, $a_{\ell m}$ and $C_{\ell}$ are the observed CMB coefficients and power spectrum, respectively, $\Braket{\cdots}_{\rm MC}$ denotes an ensemble average of Gaussian realizations, and 
\begin{eqnarray}
N^2 \equiv \sum_{\ell_1 \ell_2 \ell_3} (-1)^{\ell_1 + \ell_2 + \ell_3}
\frac{B_{\ell_1 \ell_2 \ell_3}^2}{C_{\ell_1} C_{\ell_2} C_{\ell_3}}  
\end{eqnarray}
is a normalization factor.\footnote{Note that we here express the estimator identical to ref.~\cite{Shiraishi:2014roa} with different notations.} All experimental features, such as beams, noise, galactic and point source masks, are assumed to be included in the formula above. The form of the estimator displayed here is based on the so-called diagonal covariance approach; namely, we ignore the off-diagonal components of the covariance matrix, essentially replacing $(C^{-1})_\ell$ with $1/C_{\ell}$. This approximation produces in principle suboptimal weights. However, the {\it Planck} team \cite{Ade:2013ydc} recently showed that the diagonal approximation can be adopted with minimum loss of optimality ($\sim 5 \%$) by resorting to a simple recursive inpainting pre-filtering procedure. We will adopt this approach in our present analysis.

The CMB temperature bispectrum sourced by parity-odd NG models always takes pure imaginary numbers and is characterized by the following $\ell$-space domain:
\begin{eqnarray}
\ell_1 + \ell_2 + \ell_3 = {\rm odd} ~, \ \
|\ell_1 - \ell_2| \leq \ell_3 \leq \ell_1 + \ell_2 ~.
\end{eqnarray}
It will be convenient for us to introduce a ``spin-weighted reduced bispectrum", $b_{\ell_1 \ell_2 \ell_3}$, given by 
\begin{eqnarray}
B_{\ell_1 \ell_2 \ell_3} &\equiv& h_{\ell_1 \ell_2 \ell_3}^{\{xyz\}} b_{\ell_1 \ell_2 \ell_3} ~, \\ 
h_{\ell_1 \ell_2 \ell_3}^{x~y~z} 
&\equiv& 
\sqrt{\frac{(2\ell_1 + 1)(2\ell_2 + 1)(2\ell_3 + 1)}{4\pi}}
\left(
  \begin{array}{ccc}
  \ell_1 & \ell_2 & \ell_3 \\
  x & y & z 
  \end{array}
 \right)~,
\end{eqnarray}
where the notation $\{a,b,c\}$ means permutations over the indices $a$, $b$ and $c$: $A_{\{a} A_{b} A_{c\}} \equiv \frac{1}{6}A_a A_b A_c + 5~{\rm perms~in~}a,b,c $, and we shall choose the spin set as $(x,y,z) = (1,1,-2)$ without loss of generality. In principle, the parity-odd CMB bispectrum cannot be written in separable form, since it only arises from higher spin perturbations, i.e.,  vector or tensor modes, and the $k$ dependence of the resulting primordial NG is generally tangled. For non-factorizable templates, the estimator \eqref{eq:estimator} is unfortunately characterized by a computationally unfeasible ${\cal O}(\ell_{\rm max}^5)$ scales. However, as shown below, this computational difficulty can be overcome, if such non-separable bispectrum can be approximately expanded in terms of a finite sets of (non-orthonormal) real separable basis elements (modes), which we denote as:
\begin{eqnarray}
\frac{v_{\ell_1} v_{\ell_2} v_{\ell_3}}{i \sqrt{C_{\ell_1} C_{\ell_2} C_{\ell_3}} } b_{\ell_1 \ell_2 \ell_3}
&=& 
\sum_{ijk} \alpha_{ijk}^Q Q_{ijk}(\ell_1, \ell_2, \ell_3) ~, \label{eq:Qdec} \\ 
Q_{ijk}(\ell_1, \ell_2, \ell_3) 
&\equiv& q_{\{i}(\ell_1) q_{j}(\ell_2) q_{k\}}(\ell_3) \in \mathbb{R}  ~, 
\end{eqnarray}
where $v_\ell$ is an arbitrary weight function to adjust total the $\ell$ scaling (it can be tuned in order to improve convergence). The separability of the basis bispectrum templates, and the orthonormality of the spin-weighted spherical harmonics, allow to rewrite the estimator using the expansion coefficients $\alpha_n^Q$ and $\beta_n^Q$: 
\begin{eqnarray}
{\cal E} = 
\frac{ \sum_n \alpha_{n}^Q \beta_{n}^Q}{\sum_{np} \alpha_{n}^Q \gamma_{np} \alpha_{p}^Q}
~, \label{eq:estimator_Q}
\end{eqnarray}
where $\gamma_{np} \equiv \Braket{Q_n, Q_p}_o$ denotes the inner product of the modal bases in the parity-odd domain, defined as 
\begin{eqnarray}
\Braket{f, g}_o
&\equiv& \sum_{\ell_1 + \ell_2 + \ell_3 = {\rm odd}} 
\left( \frac{h_{\ell_1 \ell_2 \ell_3}^{\{xyz\}}}{v_{\ell_1} v_{\ell_2} v_{\ell_3}} \right)^2
f(\ell_1, \ell_2, \ell_3) g(\ell_1, \ell_2, \ell_3)~,
\end{eqnarray} 
and for convenience the triples $ijk$, associated to a given template $Q$, were labeled by means of a single index $n$. Note that $\alpha_n^Q$ and $\beta_n^Q$ only take real numbers also in the parity-odd case, so that both numerator and denominator in the estimator expression \eqref{eq:estimator_Q} are real, making the measured $f_{\rm NL}$ always real valued. The $\alpha^Q$ coefficients are associated to the expansion of the theoretical bispectrum template via formula (\ref{eq:Qdec}), while the $\beta^Q$ coefficients are connected with the observed $a_{\ell m}$ through the definition:
\begin{eqnarray}
\beta_{ijk}^{Q} 
&\equiv& \frac{1}{i}
\int d^2 \hat{\bf n} 
 \left[ {}_{\{-x}M_{\{i}^{(o)} ~  {}_{-y}M_{j}^{(e)} ~ {}_{-z\}}M_{k\}}^{(e)} 
- 3 \Braket{{}_{\{-x}M_{\{i}^{(o)}~ {}_{-y}M_{j}^{(e)}}_{\rm MC} {}_{-z\}}M_{k\}}^{(e)} 
\right. \nonumber \\ 
&& \left.\qquad\quad+ 
{}_{\{-x}M_{\{i}^{(e)} ~  {}_{-y}M_{j}^{(o)} ~ {}_{-z\}}M_{k\}}^{(e)} 
- 3 \Braket{{}_{\{-x}M_{\{i}^{(e)}~ {}_{-y}M_{j}^{(o)}}_{\rm MC} {}_{-z\}}M_{k\}}^{(e)} 
\right. \nonumber \\ 
&&\left.\qquad\quad + 
{}_{\{-x}M_{\{i}^{(e)} ~  {}_{-y}M_{j}^{(e)} ~ {}_{-z\}}M_{k\}}^{(o)} 
- 3 \Braket{{}_{\{-x}M_{\{i}^{(e)}~ {}_{-y}M_{j}^{(e)}}_{\rm MC} {}_{-z\}}M_{k\}}^{(o)} 
\right. \nonumber \\ 
&&\left.\qquad\quad + 
{}_{\{-x}M_{\{i}^{(o)} ~  {}_{-y}M_{j}^{(o)} ~ {}_{-z\}}M_{k\}}^{(o)} 
- 3 \Braket{{}_{\{-x}M_{\{i}^{(o)}~ {}_{-y}M_{j}^{(o)}}_{\rm MC} {}_{-z\}}M_{k\}}^{(o)} 
\right] ~, 
\end{eqnarray} 
with the parity and spin-dependent pixel-space maps defined by 
\begin{eqnarray}
{}_x M_i^{(o/e)}(\hat{\bf n}) \equiv \sum_{\ell = {\rm odd / even}} \sum_m q_i(\ell) \frac{a_{\ell m} }{v_{\ell} \sqrt{C_\ell}}  {}_x Y_{\ell m}(\hat{\bf n}) ~.
\end{eqnarray}
The pixel-space cubic statistics above is manifestly separable and can evaluated with ${\cal O}(\ell_{\rm max}^3)$ numerical operations. Starting from the input (not necessarily factorized) bispectrum template $b_{\ell_1 \ell_2 \ell_3}$, we can extract the coefficients $\alpha_n^Q$ from 
\begin{eqnarray}
\alpha_n^Q &=& \sum_p \gamma_{np}^{-1} 
\Braket{ \frac{v_{\ell_1} v_{\ell_2} v_{\ell_3} b_{\ell_1 \ell_2 \ell_3}}{i \sqrt{C_{\ell_1} C_{\ell_2} C_{\ell_3}}}, Q_p(\ell_1,\ell_2,\ell_3) }_o ~. \label{eq:alphaQ}
\end{eqnarray}
Correlating the vector $\alpha_n$ (theory) with the vector $\beta_n$ (observation) we arrive at computationally feasible implementation of an estimator for $f_{\rm NL}$.

Although $Q_n$ is a complete but not orthonormal basis, we can also generate a complete orthonormal basis by means of a suitable rotation in bispectrum space. The orthonormal basis templates will be called $R_n$, and they are given by  $R_n = \sum_{p} \lambda_{np} Q_p$, where $\lambda$ is a lower triangular matrix implicitly defined by the condition $\gamma^{-1} = \lambda^\top \lambda$. The bispectrum can be expanded as a linear combination of $R_n$: 
\begin{eqnarray}
\frac{v_{\ell_1} v_{\ell_2} v_{\ell_3}}{i \sqrt{C_{\ell_1} C_{\ell_2} C_{\ell_3}} } b_{\ell_1 \ell_2 \ell_3} &=& \sum_n \alpha_n^R R_n(\ell_1,\ell_2,\ell_3) ~, \\
\alpha_n^R &=& 
\Braket{ \frac{v_{\ell_1} v_{\ell_2} v_{\ell_3} b_{\ell_1 \ell_2 \ell_3}}{i \sqrt{C_{\ell_1} C_{\ell_2} C_{\ell_3}}}, R_n(\ell_1,\ell_2,\ell_3) }_o ~,
\end{eqnarray}
and we obtain  
\begin{eqnarray}
{\cal E} = 
\frac{\sum_n \alpha_{n}^R \beta_{n}^R}{\sum_{n} (\alpha_{n}^R)^2}
~. \label{eq:estimator_R}
\end{eqnarray}
For consistency checks, both estimator forms \eqref{eq:estimator_Q} and \eqref{eq:estimator_R} are applied in the following bispectrum analyses. 

By use of the above equations, we can also easily show that the ensemble average of $\beta_n^R$ of the NG maps sourced by a theoretical bispectrum is given by $\alpha_n^R$:
\begin{eqnarray}
\Braket{\beta_n^R} = f_{\rm NL} \alpha_n^R~. \label{eq:beta_ave}
\end{eqnarray}
Hence, if $\beta_n^R$ can be accurately extracted from particular observational data, we can reconstruct the parity-odd CMB bispectrum from the data with
\begin{eqnarray}
b_{\ell_1 \ell_2 \ell_3}^{\rm obs} = \frac{i \sqrt{C_{\ell_1} C_{\ell_2} C_{\ell_3}}}{v_{\ell_1} v_{\ell_2} v_{\ell_3}} \sum_n \beta_n^R R_n(\ell_1, \ell_2 ,\ell_3)~. 
\label{eq:bis_rec}
\end{eqnarray}
Of course the fact that we are truncating the sum above to a finite number of templates implies that we are producing a smoothed reconstruction of the actual data bispectrum.
We will perform this bispectrum reconstruction from WMAP data in section~\ref{sec:reconstruction}.

\subsection{Numerical tests with simulated non-Gaussian maps}

\begin{table}[t]
\begin{center}
  \begin{tabular}{|c||c|c|c|c|c|c|c|c|} \hline
    & ideal: $f_{\rm sky} = 1$ & WMAP: $f_{\rm sky} = 0.688$ \\ \hline
    average & 5.09 & 5.16 \\ \hline
    $1\sigma$ error & 0.89 & 1.08 \\  \hline
  \end{tabular}
\end{center}
\caption{
Average values of $f_{\rm NL}^P \times 10^{-4}$ estimated from 100 simulated NG maps with an input value: $f_{\rm NL}^P = 5 \times 10^4$ in the pseudoscalar model, and the $1\sigma$ errors estimated from 1000 simulated Gaussian maps, assuming the ideal full-sky noiseless and WMAP-like experiments. The average values recover the input value $5 \times 10^4$ and the error bars are almost identical to the values expected in the Fisher matrix forecast, namely $\delta f_{\rm NL} = 1/\sqrt{f_{\rm sky} F} = 0.90 \times 10^{4}$ (ideal) and $1.08 \times 10^{4}$ (WMAP), respectively.} 
\label{tab:pseudo_sim}
\end{table}

\begin{figure}[t]
\begin{tabular}{c}
  \begin{minipage}{1\hsize}
  \begin{center}
    \includegraphics[width = 0.85\textwidth]{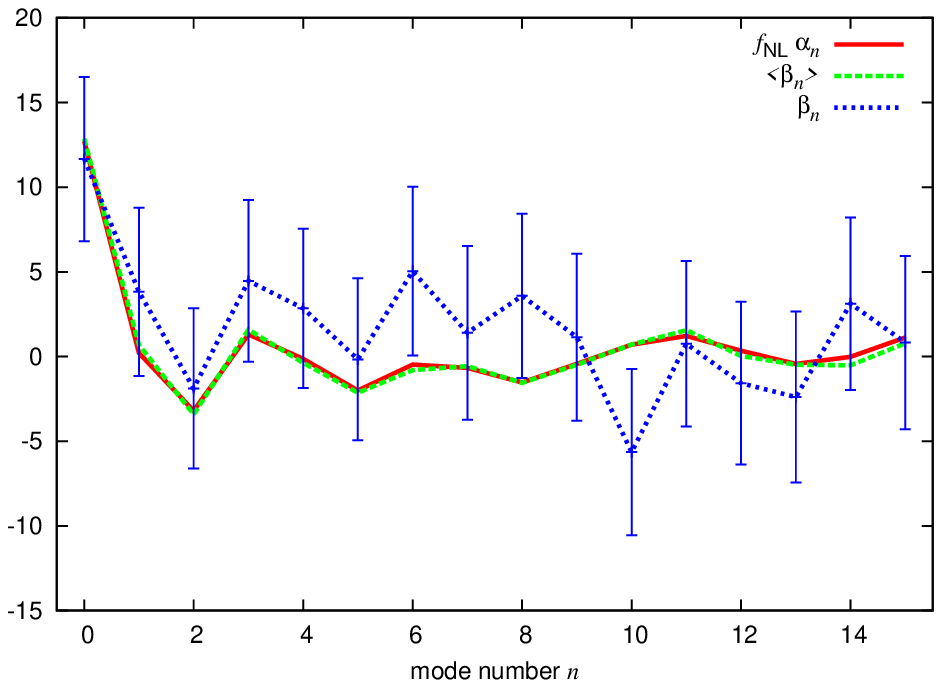}
  \end{center}
\end{minipage}
\end{tabular}
\\ 
\begin{tabular}{c}
  \begin{minipage}{1\hsize}
  \begin{center}
    \includegraphics[width = 0.85\textwidth]{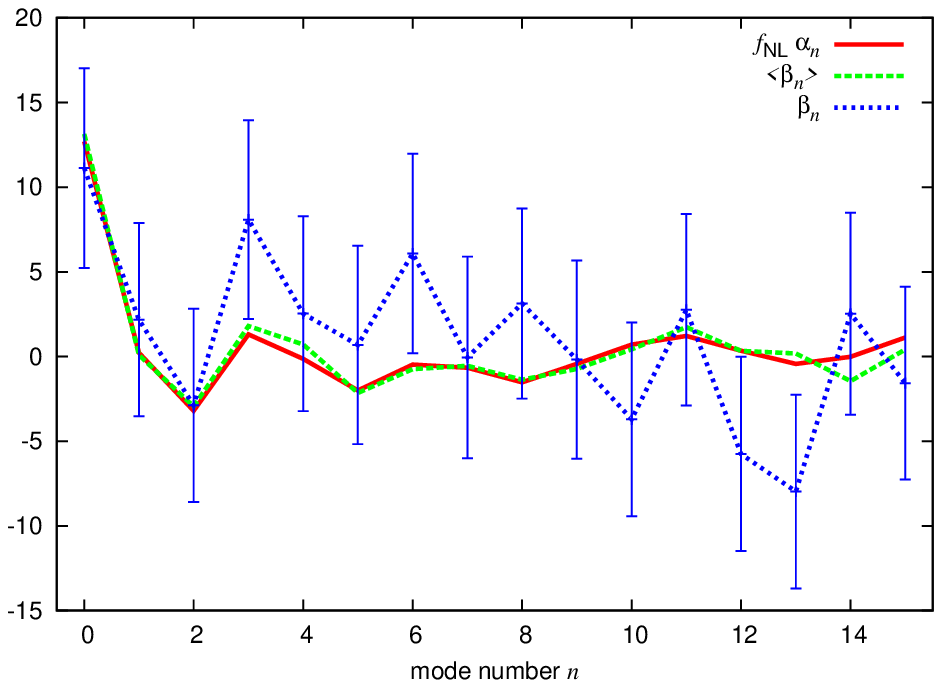}
  \end{center}
\end{minipage}
\end{tabular}
  \caption{
Average values of $\beta_n^R$ obtained from 100 simulated NG maps in the pseudoscalar model with $f_{\rm NL}^P = 5 \times 10^4$, namely $\Braket{\beta_n^R}$, and a particular $\beta_n^R$ of them with the $2\sigma$ error bars in the ideal (top panel) and WMAP-like (bottom panel) analyses. For reproducibility check, we here also plot the input modal coefficients, namely $f_{\rm NL}^P \alpha_n^R$. Note that each curve in the ideal and WMAP-like cases are sourced by the identical random Gaussian seeds each other. Here we pick up first 15 modal coefficients.
}
\label{fig:betaR_XXX5e18}
\end{figure}

Before moving to actual data analysis, we checked the validity of our estimators and numerical approaches by estimating $f_{\rm NL}$ from realistic simulated NG maps with known $f_{\rm NL}$. Since in this work we applied our pipeline to the analysis of WMAP 9-year data, our simulation were also produced at WMAP resolution, taking $\ell_{\rm max} = 500$ and HEALPix $n_{\rm side} = 512$ corresponding to ${\cal O}(10^6)$ pixels in the maps. 

We considered tensor-mode nonlinearity parameter in the pseudoscalar model, and measured ($f_{\rm NL}^P$, defined in section~\ref{sec:constraints}), from 100 simulated NG maps. We chose an input $f_{\rm NL}^P$ of $5 \times 10^4$, corresponding, for the model under examination, to an expected NG detection, for a WMAP-like experiment, at $\sim 5 \sigma$ significance. The parity-odd NG maps were produced using the fast separable modal-based algorithm discussed in ref.~\cite{Shiraishi:2014roa}. 
The tests were performed both on ``ideal" full-sky noiseless NG maps and on more realistic simulation including WMAP instrumental features \cite{Bennett:2012zja, Hinshaw:2012aka}. More specifically, we accounted for partial-sky coverage, using the KQ75 mask (recommended by the WMAP team for NG studies of WMAP data), and included an anisotropic noise component. As for the actual data analysis, we inpainted masked regions using the simple recursive inpainting technique described in ref.~\cite{Ade:2013ydc}. Error estimation is done Monte Carlo, using 1000 Gaussian simulations both for the ``ideal" and ``realistic" case. The Monte Carlo averaging for linear term computation is performed on 500 Gaussian realizations. For more technical details on our analysis we refer the reader to section~\ref{sec:reconstruction}. 

In place of the usual $v_\ell = (2\ell + 1)^{1/6}$ weighting, generally adopted for parity-even studies, we here choose $v_\ell = (2\ell + 1)^{1/6} (C_\ell^{\rm scal} / C_\ell^{\rm tens})^{1/2}$, with $C_\ell^{\rm scal}$ and $C_\ell^{\rm tens}$ denoting theoretical scalar-mode and tensor-mode temperature power spectra, respectively. This choice achieves faster convergence of the modal decomposition when tensor bispectra are involved. 
With this choice, we see that convergence (for parity-odd models), at WMAP angular resolution, is achieved with 30 basis templates, 
while the old $v_\ell = (2\ell + 1)^{1/6}$ weighting needs more than 200 basis templates. Our basis templates are composed of polynomial eigenfunctions and a special mode function enhanced at the squeezed limit (located at $n=1$), which have been used in the parity-even bispectrum analysis by the {\it Planck} team \cite{Ade:2013ydc}.

Table~\ref{tab:pseudo_sim} describes the results of our validation tests. We report recovered mean values and $1\sigma$ errors on $f_{\rm NL}^P$, both for the all-sky noise-free ideal case, and for WMAP-like simulations. We thus confirm, 
according to theoretical expectations,
that our parity-odd $f_{\rm NL}$ estimator is both optimal and unbiased. 

Figure~\ref{fig:betaR_XXX5e18} plots the mean values of 100 $\beta_n^R$ realizations ($\Braket{\beta_n^R}$), a particular $\beta_n^R$ for a single realization (with $2\sigma$ error bars), and the input modal coefficients $f_{\rm NL}^P \alpha_n^R$ given by the theoretical bispectrum in the ideal and WMAP-like cases. For comparison between the ideal and WMAP-like test, simulations in the two cases have  identical  Gaussian random seeds. It is obvious that, for both cases, $\Braket{\beta_n^R}$ match well, within error bars, the theoretical  $f_{\rm NL}^P \alpha_n^R$ coefficients, as expected from eq.~\eqref{eq:beta_ave}. Again as expected, instrumental features (beam shape, inhomogeneous noise, sky cut) change the $\beta_n^R$ spectra and broaden the errors (see blue curves). 

The above results make us confident in the accuracy of our analytic derivation and numerical treatment.

\section{Reconstructed WMAP parity-odd bispectrum}\label{sec:reconstruction}

Having validated our pipeline, we are now ready to extract parity-odd bispectrum information from WMAP temperature data by means of the separable modal methodology. 


Our input map is obtained by coadding the V and W band foreground reduced WMAP9 temperature maps~\cite{Bennett:2012zja, Hinshaw:2012aka}. As a cross-check, we will also run our pipeline on raw maps. As seen in the next section, the $f_{\rm NL}$ constraints obtained in the two cases are very close, with central values differing by no more than  $\sigma/2$ in the most discrepant case. This suggests, on one hand, that NG contribution from foreground contamination in the raw map, if present, is already small and, on the other hand, that the foreground cleaning procedure did not generate spurious NG from e.g. oversubtraction the foreground templates. Beams and hit count maps for the different channels were obtained from the Lambda website \cite{Lambda}. As also done for the validation phase, we adopt the KQ75 mask ($f_{\rm sky} = 0.688$) and fill masked regions using the recursive inpainting procedure described in ref.~\cite{Ade:2013ydc}.
 
As a further validity check, besides the parity-odd shapes of interest, we also constrain standard parity-even NGs of the local, equilateral and orthogonal types. This allows us to carry out a successful consistency check between our results and the output of the analysis carried out by the WMAP team for the same shape. We then find that we are in very good agreement with results from the WMAP team~\cite{Komatsu:2008hk, Komatsu:2010fb, Bennett:2012zja}


\begin{figure}[t]
\begin{tabular}{c}
    \begin{minipage}{1.0\hsize}
  \begin{center}
    \includegraphics[width = 1\textwidth]{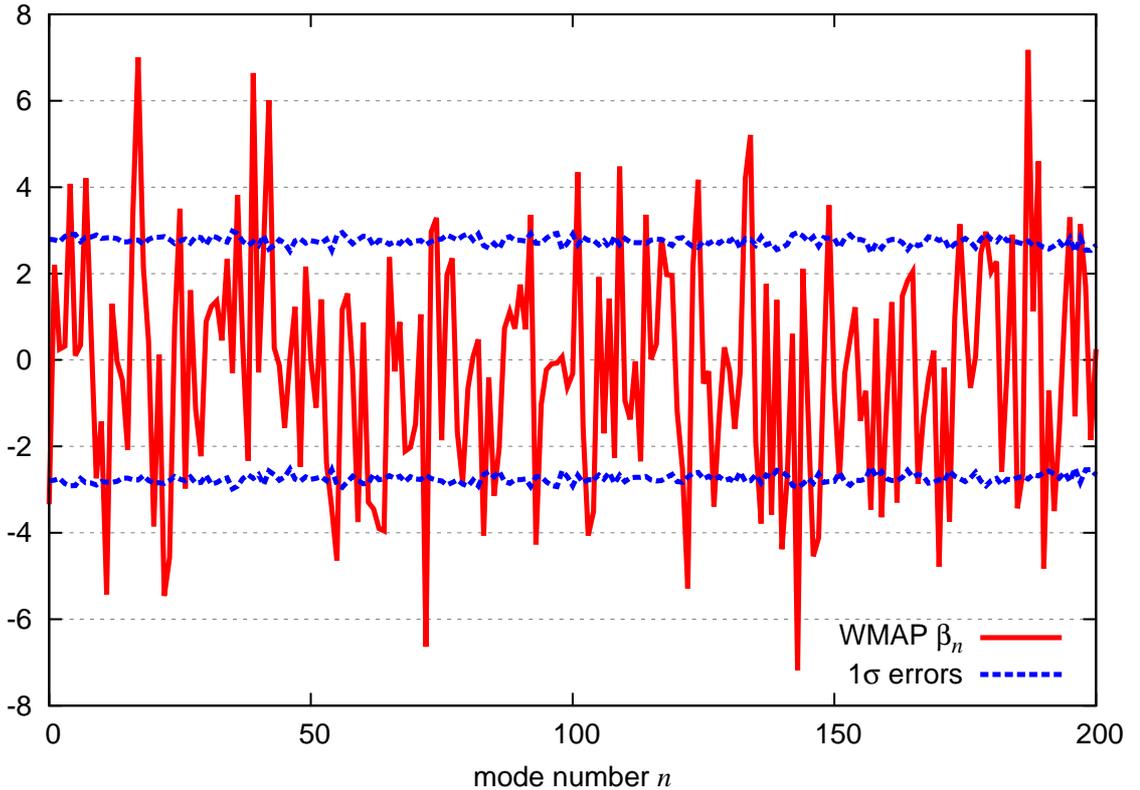}
  \end{center}
\end{minipage}
\end{tabular}
  \caption{Recovered modal coefficients $\beta_n^R$ from the coadded V+W WMAP data. The $1\sigma$ errors are estimated from 500 simulated Gaussian maps. As expected, they are close to a theoretically expected value: $\sqrt{6/f_{\rm sky}} \simeq 3$.}
  \label{fig:betaR_WMAP}
\end{figure}

The first 200 modes of the $\beta^R$ coefficients extracted from WMAP temperature data are displayed in figure~\ref{fig:betaR_WMAP}. The variances are estimated from 500 Gaussian realization processed as described in the previous section. Note that, 
being at this stage interested in bispectrum reconstruction, rather than $f_{\rm NL}$ estimation, here we decide to compute $\beta_n^R$ using the standard $v_\ell = (2\ell + 1)^{1/6}$ weighting. We find that this makes the estimation of $\beta_n^R$ more stable for high modal numbers  ($n > 50$), with respect to the alternative $v_\ell = (2\ell + 1)^{1/6} (C_\ell^{\rm scal} / C_\ell^{\rm tens})^{1/2}$ ``parity-odd weighting"  (as explained above, $\alpha_n^R$ converges more slowly when expanding tensor shapes with these weights; however we are not concerned with this issue at this stage, as we are preforming a model independent analysis. Conversely, we are not worried about instabilities at $n>50$ with the parity-odd weighting, as we use this approach for $f_{\rm NL}$ estimation, but in that case we need only $30$ modes to reconstruct accurately the theoretical shapes of interest and fit to the data). It is apparent from the figure that the variance of each $\beta_n^R$ properly converges to its theoretically expected value, that is  $\sqrt{\Braket{(\beta_n^R)^2}} \to \sqrt{6/f_{\rm sky}} \simeq 3$. Most of our $\beta_n^R$ are totally consistent with $0$. This is already telling us that parity-odd shapes are not going to be detected at a high degree of significance, as it will be explicitly shown in the next section. Let us emphasize again that our  $\beta_n^R$ are completely model-independent observables. We can thus use the above results to obtain WMAP limits on {\em all} types of parity-odd NGs, and not just on the shapes analyzed in the next section (provided the additional 
parity-odd templates can be expanded accurately using the $200$ modes constrained here).

\begin{figure}[t]
\begin{tabular}{c}
    \begin{minipage}{1.0\hsize}
  \begin{center}
    \includegraphics[width = 1\textwidth]{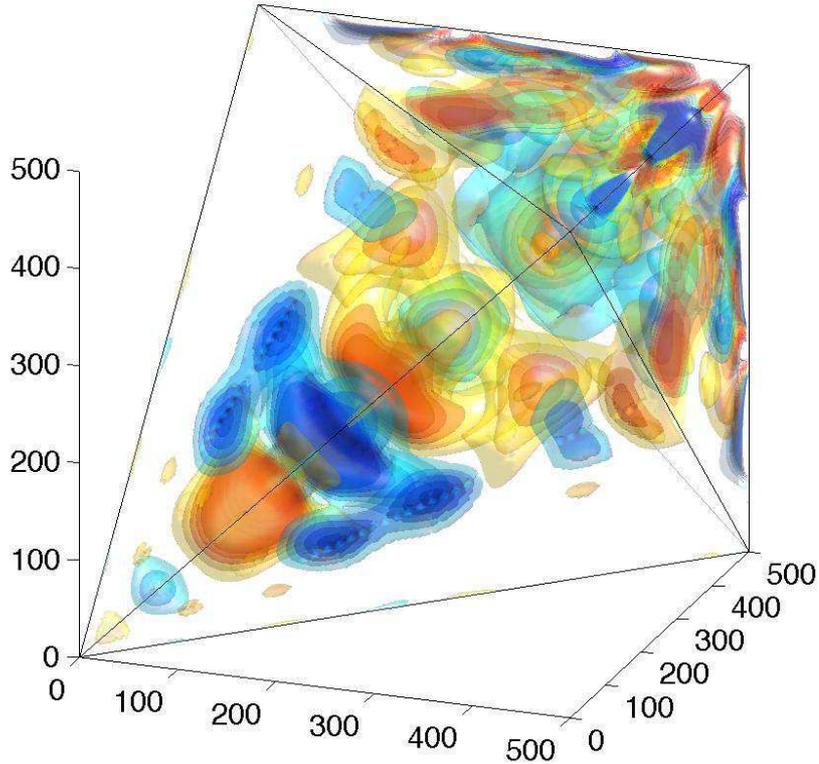}
  \end{center}
\end{minipage}
\end{tabular}
  \caption{3D representation of the parity-odd CMB bispectrum ($\ell_1 + \ell_2 + \ell_3 = {\rm odd}$) recovered from the WMAP temperature data estimated with eq.~\eqref{eq:bis_rec}.}
  \label{fig:WMAP_odd_bispectrum}
\end{figure}

A straightforward substitution of the measured $\beta_n^R$  into eq.~\eqref{eq:bis_rec} allows to reconstruct the observed parity-odd part of the WMAP bispectrum. This is shown in figure~\ref{fig:WMAP_odd_bispectrum}. Of course, as already noted for the $\beta_n^R$, the bispectrum configurations are well consistent with cosmic variance and noise fluctuations. It is anyway visually apparent that all the highest values are distributed around the equilateral configurations $\ell_1 \approx \ell_2 \approx \ell_3$. This, when fitting specific NG templates to the data, will produce a slight preference for parity-odd equilateral NG with respect to parity-odd squeezed NG, as shown in the next section (although in all cases we will be well consistent with 
Gaussianity within error bars).

\section{Constraints on parity-odd non-Gaussianities}\label{sec:constraints}

\begin{figure}[t]
\begin{tabular}{c}
    \begin{minipage}{1.0\hsize}
  \begin{center}
    \includegraphics[width = 1\textwidth]{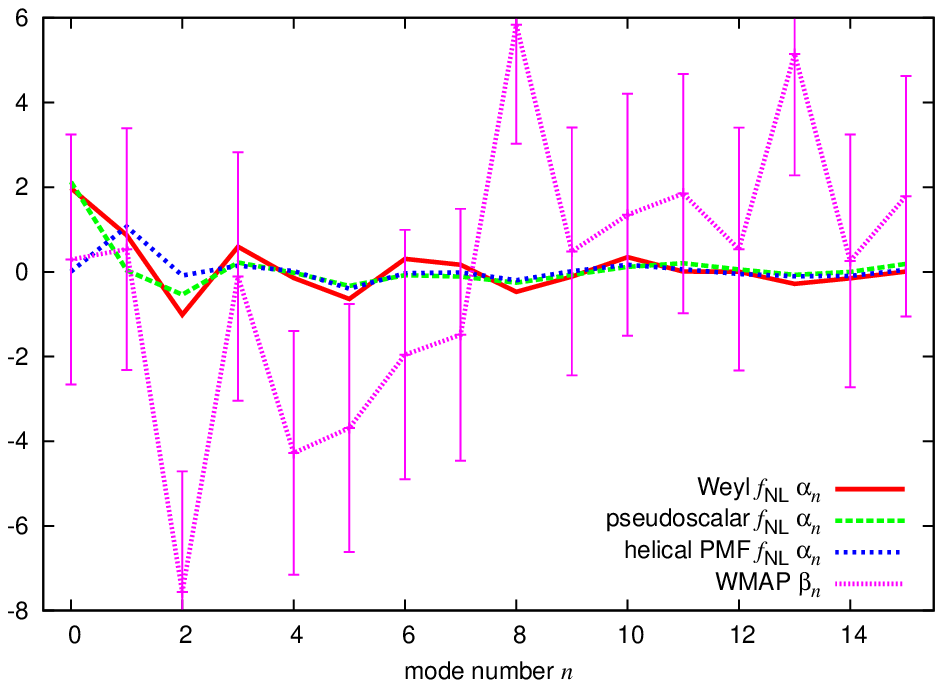}
  \end{center}
\end{minipage}
\end{tabular}
  \caption{Comparison of the theoretical modal coefficients $f_{\rm NL} \alpha_n^R$ and the WMAP $\beta_n^R$ with the $1\sigma$ error bars. As $f_{\rm NL}$, we adopt the best-fit values constrained from the WMAP data. The $n = 0$ mode expresses the constant component, while the $n=1$ mode is sensitive to the squeezed-limit signals. Here we present the first 15 modes.}
  \label{fig:alphaR_betaR}
\end{figure}

In this section, we estimate the WMAP constraints on the parity-odd NGs predicted in three Early Universe models, including the Weyl dual cubic action \cite{Maldacena:2011nz, Soda:2011am, Shiraishi:2011st}, a rolling pseudoscalar \cite{Cook:2013xea, Shiraishi:2013kxa}, and the helical primordial magnetic field (PMF) model \cite{Shiraishi:2012sn}. 

In such theories, sizable primordial tensor bispectra can be created. These are written in the form:
\begin{eqnarray}
\Braket{\prod_{i=1}^3 h_{{\bf k}_i}^{(\lambda_i)}} &=&  (2\pi)^3 
\delta^{(3)} \left( \sum_{i=1}^3 {\bf k}_i \right) B_{ {\bf k}_1 {\bf k}_2 {\bf k}_3}^{\lambda_1 \lambda_2 \lambda_3} ~, 
\end{eqnarray}
where $h_{\bf k}^{(\lambda)}$ is the gravitational wave on the spin $\lambda = \pm 2$ state, defined in $\delta g_{ij}^{TT} / a^2 = \int \frac{d^3 {\bf k}}{(2\pi)^3} \sum_{\lambda = \pm 2} h_{\bf k}^{(\lambda)}e_{ij}^{(\lambda)}(\hat{\bf k}) e^{i {\bf k} \cdot {\bf x}}$, with $e_{ij}^{(\lambda)}(\hat{\bf k})$ denoting a transverse and traceless polarization tensor obeying $e_{ij}^{(\lambda)}(\hat{\bf k}) e_{ij}^{(\lambda')}(\hat{\bf k}) = 2 \delta_{\lambda, -\lambda'}$ and $e_{ij}^{(\lambda) *}(\hat{\bf k}) = e_{ij}^{(-\lambda)}(\hat{\bf k}) = e_{ij}^{(\lambda)}(- \hat{\bf k})$ \cite{Shiraishi:2010kd}. Parity violation arises when the equality between $B^{\lambda_1 \lambda_2 \lambda_3}$ and their spin-flipped versions breaks (e.g. $B^{+2 +2 +2 } \neq B^{-2 -2 -2 }$). This parity-odd information is directly projected into the parity-odd $\ell_1 + \ell_2 + \ell_3$ domain of the CMB temperature bispectrum via harmonic transforms, as shown in \cite{Shiraishi:2010sm, Shiraishi:2010kd}, yielding:
\begin{eqnarray}
\Braket{\prod_{i=1}^3 a_{\ell_i m_i}} 
&=& \left[ \prod_{i=1}^3 i^{\ell_i} 
\int \frac{d^3 {\bf k}_i}{2 \pi^2} {\cal T}_{\ell_i}(k_i) 
\sum_{\lambda_i = \pm 2} 
{}_{-\lambda_i}Y_{\ell_i m_i}^*(\hat{\bf k}_i)
\right]
\Braket{\prod_{i=1}^3 h_{{\bf k}_i}^{(\lambda_i)} } \nonumber \\ 
&\equiv& f_{\rm NL} B_{\ell_1 \ell_2 \ell_3} 
\left(
\begin{array}{ccc}
\ell_1 & \ell_2 & \ell_3 \\
m_1 & m_2 & m_3
\end{array}
\right)
~,
\end{eqnarray}
where ${\cal T}_{\ell}(k)$ is the radiation transfer function of the tensor temperature mode. Such temperature bispectra are only enhanced on large scales by the Integrated Sachs-Wolfe (ISW) effect and hence the signal-to-noise ratios are saturated for $\ell \gtrsim 100$ \cite{Shiraishi:2014roa}. 

To parametrize the amplitude of such tensor bispectra, we shall introduce two types of $f_{\rm NL}$, using the squeezed- and equilateral-limit values of the $\lambda_i = 2$ bispectrum component, and normalizing to the corresponding parity-even shapes:  
\begin{eqnarray}
f_{\rm NL}^{\rm sq} &\equiv& \lim_{ \substack{k_1 \to k_2 \\ k_3 \to 0}} 
\frac{B_{ {\bf k}_1 {\bf k}_2 {\bf k}_3}^{+2 +2 +2}}{B_{k_1 k_2 k_3}^{\zeta \rm loc}(f_{\rm NL}^{\zeta \rm loc} = 1)}
~, \label{eq:fnl_def_sq} \\
f_{\rm NL}^{\rm eq} &\equiv& \lim_{k_i \to k}
\frac{B_{ {\bf k}_1 {\bf k}_2 {\bf k}_3}^{+2 +2 +2}}{B_{k_1 k_2 k_3}^{\zeta \rm eq}(f_{\rm NL}^{\zeta \rm eq} = 1)} ~, \label{eq:fnl_def_eq}
\end{eqnarray}
where, as already mentioned, the normalizing factors at denominator are the usual local ($B_{k_1 k_2 k_3}^{\zeta \rm loc}$) and equilateral ($B_{k_1 k_2 k_3}^{\zeta \rm eq}$) bispectra of curvature perturbations; in other words, the $f_{\rm NL}$ explicitly appearing in the formulae above are the usual nonlinearity parameters for scalar NGs $f_{\rm NL}^{\zeta \rm loc}$ and $f_{\rm NL}^{\zeta \rm eq}$.

In the following section, we consider three models giving rise to parity-odd NG described by the above ansatz. We estimate $f_{\rm NL}$ for each model, following the methodology summarized in section~\ref{sec:modal}. Contrarily to what done in the previous section, we now use the parity-odd $v_\ell = (2\ell + 1)^{1/6} (C_\ell^{\rm scal} / C_\ell^{\rm tens})^{1/2}$ weighting. In this way we achieve rapid convergence of the $\alpha_n^R$ theoretical expansions, allowing us to obtain 
accurate $f_{\rm NL}$ estimates with only $30$ modes; as mentioned in the previous section, the drawback of this weight choice is to produce some numerical instabilities in $\beta_n$ estimation; however this happens only at high $n$, $n>50$, hence it is not of any concern here. For error bars estimation, and to compute linear terms, we use our usual sets of 1000 and 500 inpainted Gaussian maps respectively. 
The observed $\beta_n^R$ and the theoretical coefficients $f_{\rm NL} \alpha_n^R$ for the three models are depicted in figure~\ref{fig:alphaR_betaR}. We are now going to describe our results model by model. 

\subsection{Weyl model}

Parity violation of graviton NGs was firstly discussed in the framework of  Weyl gravity \cite{Maldacena:2011nz, Soda:2011am, Shiraishi:2011st}.  Here we focus on signatures of a dual cubic action with time-dependent coupling~\cite{Shiraishi:2011st}
\begin{eqnarray}
S = \int d\tau d^3 x \frac{f(\tau)}{\Lambda^2} \widetilde{W}^{\alpha \beta}{}_{\gamma \delta} W^{\gamma \delta}{}_{\sigma \rho} W^{\sigma \rho}{}_{\alpha \beta} ~, \label{eq:action}
\end{eqnarray}
where $ W_{\mu \nu \gamma \delta}$ and $ \widetilde{W}_{\mu \nu \gamma \delta}$ is the Weyl tensor and its dual, respectively, and $\Lambda$ determines a energy scale of this action. A running coupling $f(\tau)$ is simply assumed as a power-law function of the conformal time, namely $f(\tau) = (\tau / \tau_*)^n$, with a pivot scale $\tau_* = - 14 ~{\rm Gpc}$. 

An explicit form of the gravitational wave bispectrum is \cite{Shiraishi:2011st}
\begin{eqnarray}
B_{ {\bf k}_1 {\bf k}_2 {\bf k}_3}^{\lambda_1 \lambda_2 \lambda_3} &=& 
8 \left( \frac{H}{M_{\rm pl}} \right)^6 \left( \frac{H}{\Lambda} \right)^2  
\sin\left(\frac{n \pi}{2}\right) \frac{\Gamma(6 + n)}{ k_t^{6} (- k_t \tau_*)^{n}} \nonumber \\ 
&&\times 
i \eta_{ijk} 
\left[ e_{kq}^{(-\lambda_1)}(\hat{\bf k}_1) 
\left\{- 3 e_{jm}^{(-\lambda_2)}(\hat{\bf k}_2)  e_{iq}^{(-\lambda_3)}(\hat{\bf k}_3) 
\hat{k}_{3m} 
+ e_{mi}^{(-\lambda_2)}(\hat{\bf k}_2)  e_{mq}^{(-\lambda_3)}(\hat{\bf k}_3)  \hat{k}_{3j} \right\} \right. \nonumber \\
&&\qquad\quad\left. 
+ e_{pj}^{(-\lambda_1)}(\hat{\bf k}_1)  e_{pm}^{(-\lambda_2)}(\hat{\bf k}_2) 
\hat{k}_{1k} \hat{k}_{2l}
\left\{ e_{il}^{(-\lambda_3)}(\hat{\bf k}_3)  \hat{k}_{3m} 
- e_{im}^{(-\lambda_3)}(\hat{\bf k}_3) \hat{k}_{3l} \right\}  \right] \nonumber \\ 
&& + 5 \ {\rm perms \ in} \ ({\bf k}_1, \lambda_1), ({\bf k}_2, \lambda_2), ({\bf k}_3, \lambda_3)~,
 \label{eq:bis_Weyl}
\end{eqnarray}
where $H$ is the Hubble parameter during inflation, $M_{\rm pl} \equiv (8\pi G)^{-1/2}$ is the reduced Planck mass, $\eta_{ijk}$ is a 3D antisymmetric tensor and $k_t \equiv k_1 + k_2 + k_3$. Interestingly, owing to the sine function, the parity-odd bispectrum vanishes for even $n$. In this paper we consider the $n = 1$ case. The corresponding bispectrum is then maximized in the equilateral limit, due to the $k_t^{-6-n}$ dependence \cite{Shiraishi:2011st}. The amplitude of the bispectrum is determined by two free parameters, $H$ and $\Lambda$. Computing the equilateral-type normalization \eqref{eq:fnl_def_eq} and translating $H$ into the tensor-to-scalar ratio $r$, we can introduce the following nonlinearity parameter:
\begin{eqnarray}
f_{\rm NL}^{W} \equiv  3 \times 10^{-19} 
 \left( \frac{M_{\rm pl}}{\Lambda} \right)^2 r^4 ~.
\end{eqnarray}
 The $1\sigma$ error on $f_{\rm NL}^{W}$ expected in a noiseless full-sky measurement is $\delta f_{\rm NL}^{W} = 1.4 \times 10^2$ \cite{Shiraishi:2014roa}.

The $\alpha^R$ spectrum arising from this template is shown in figure~\ref{fig:alphaR_betaR}. Partly as a consequence of this bispectrum peaking in the equilateral limit, the $n=0$ mode, that is the constant mode, gives the largest contribution. We also see that $\alpha_n^R$ rapidly goes to $0$ as the mode number $n$ increases, ensuring good convergence. 

The estimator sum given by the products of $\alpha_n$ and $\beta_n$ (\eqref{eq:estimator_Q} or \eqref{eq:estimator_R}) yields the following constraint on $f_{\rm NL}^W$, with measured central value at around the edge of the $1 \sigma$ confidence interval:
\begin{eqnarray}
f_{\rm NL}^W = (1.5 \pm 1.6) \times 10^2 \ \ (68\%{\rm CL})~.
\end{eqnarray}
The central value is a bit smaller than the constraint estimated from the not foreground reduced raw maps: $f_{\rm NL}^W = (2.2 \pm 1.6) \times 10^2$.

\subsection{Pseudoscalar model}

A large tensor equilateral bispectrum can also be produced in an inflationary model where the inflaton $\phi$ couples to a pseudoscalar $\chi$, reading \cite{Barnaby:2012xt}
\begin{eqnarray}
 {\cal L} =  
-\frac{1}{2} (\partial \phi)^2 - V(\phi) - \frac{1}{2} (\partial \chi)^2 - U(\chi) - \frac{1}{4}F_{\mu \nu} F^{\mu \nu} - \frac{\chi}{4f} F_{\mu \nu} \widetilde{F}^{\mu \nu} ~, \label{eq:pseudo_action}
\end{eqnarray}
where $F_{\mu \nu} \equiv \partial_\mu A_\nu - \partial_\nu A_\mu$ and $\widetilde{F}^{\mu \nu}$ are the field strength of the U(1) gauge field $A_{\mu}$ and its dual, respectively, and $f$ is a coupling constant like an axion decay constant. In this scenario, the polarized gauge fields sourced by the dual term generate the parity-violating NG gravitational waves because of the quadratic dependence of the energy momentum tensor on Gaussian gauge fields. Then, interestingly, the resulting primordial tensor bispectrum can be larger than the scalar one \cite{Cook:2013xea}; hence the CMB tensor bispectrum becomes the main NG observable in this model.

The gravitational wave bispectrum is expressed as \cite{Shiraishi:2013kxa}
\begin{eqnarray}
B^{\lambda_1 \lambda_2 \lambda_3}_{{\bf k}_1 {\bf k}_2 {\bf k}_3}  
&\approx& 4.3 \times 10^{-3} {\cal P}^3 X^3 S^{\rm eq}_{k_1 k_2 k_3} 
e_{ij}^{(-\lambda_1)}(\hat{\bf k}_1)
e_{jk}^{(-\lambda_2)}(\hat{\bf k}_2)
e_{ki}^{(-\lambda_3)}(\hat{\bf k}_3) 
\delta_{\lambda_1, 2} \delta_{\lambda_2, 2} \delta_{\lambda_3, 2}
~, \label{eq:bis_pseudo}  
\end{eqnarray}
where $S^{\rm eq}_{k_1 k_2 k_3} 
= -(k_1^{-3} k_2^{-3} + 2~{\rm perms}) - 2 k_1^{-2} k_2^{-2} k_3^{-2} 
+ (k_1^{-1} k_2^{-2} k_3^{-3} + 5~{\rm perms})$ is the usual equilateral template, ${\cal P} \approx 2.5 \times 10^{-9}$ is the scalar power spectrum and $X \equiv \epsilon \frac{e^{2\pi\xi}}{\xi^3}$ is given by a slow-roll parameter for the inflaton $\epsilon$ and a rolling parameter for the pseudoscalar $\xi \equiv \frac{\partial_t \chi}{2fH}$. Due to the $\lambda_i = +2$ polarized nature in eq.~\eqref{eq:bis_pseudo}, the resulting CMB temperature bispectrum has parity-odd signals. It is amplified in the equilateral limit ($\ell_1 \approx \ell_2 \approx \ell_3$) because of $S_{k_1 k_2 k_2}^{\rm eq}$ dependence in eq.~\eqref{eq:bis_pseudo}. Following the equilateral-type normalization \eqref{eq:fnl_def_eq}, let us define a nonlinearity parameter:
\begin{eqnarray}
f_{\rm NL}^{P} 
\equiv 10^{-14} X^3
\end{eqnarray}
The expected $1\sigma$ error given by the noiseless full-sky fisher forecast is $\delta f_{\rm NL}^P = 0.9 \times 10^4$ \cite{Shiraishi:2013kxa}.

As shown in figure~\ref{fig:alphaR_betaR}, the resulting $\alpha^R$ spectrum resembles the Weyl counterpart closely. The ratio between the constant $n=0$ mode and the squeezed $n=1$ mode is larger, compared with the Weyl case, because of the the pseudoscalar bispectrum is even more peaked on equilateral configurations than the Weyl one already was. Also in this case, we find consistency with Gaussianity:
\begin{eqnarray}
f_{\rm NL}^P = (0.8 \pm 1.1) \times 10^4 \ \ (68\% {\rm CL}) ~.
\end{eqnarray}
In the raw map analysis, we obtain again a fully consistent central value: $f_{\rm NL}^P = (1.1 \pm 1.1) \times 10^4$.

\subsection{Helical primordial magnetic field model}

If PMFs (scaling like $B_i \propto a^{-2}$) spread beyond superhorizon scales in the deeply radiation dominated era, prior to neutrino decoupling, their anisotropic stress fluctuations generate gravitational waves, reading 
\begin{eqnarray}
h_{\bf k}^{(\lambda)} \approx - 1.8 
\frac{\ln(T_B / T_\nu )}{4\pi \rho_{\gamma,0}} 
e_{ij}^{(-\lambda)}(\hat{\bf k})  \int \frac{d^3 {\bf k'}}{(2 \pi)^3} B_i({\bf k'}) B_j({\bf k} - {\bf k'}) ~, \label{eq:GW_helical}
\end{eqnarray}
where $\rho_{\gamma,0}$ is the present photon energy density, and $T_B$ and $T_\nu \simeq 1$~MeV are energy scales of the PMF generation and neutrino decoupling, respectively. Assuming stochastic creation of PMFs, the gravitational waves become highly NG (chi-square) fields because of the quadratic dependence on $B_i$. The PMF power spectrum may be generally parametrized by \cite{Caprini:2003vc}
\begin{eqnarray}
\braket{B_i ({\bf k}) B_j ({\bf k'})} &=& \frac{(2 \pi)^3}{2}
 \left[ P_B(k) (\delta_{ij} - \hat{k}_i \hat{k}_j) + i \eta_{ijk} \hat{k}_k P_{{\cal B}}(k) \right]
\delta^{(3)}({\bf k} + {\bf k'}) ~,
\end{eqnarray}
where 
\begin{eqnarray}
P_B(k) \equiv \frac{(2 \pi)^{n_B + 5} B^2_{1}}{\Gamma \left(\frac{n_B+3}{2}\right) (\frac{2\pi}{1 ~ \rm Mpc})^{n_B + 3}} k^{n_B} ~, \ \ 
P_{\cal B}(k) \equiv \frac{(2 \pi)^{n_{\cal B} + 5} {\cal B}^2_{1}}{\Gamma
 \left(\frac{n_{\cal B} + 4}{2}\right) (\frac{2\pi}{1 ~ \rm Mpc})^{n_{\cal B} + 3}} k^{n_{\cal B}} ~,
\end{eqnarray}
are the power spectra of the non-helical and helical PMFs normalized at 1 Mpc, respectively.  For a nearly scale-invariant case: $n_B = n_{\cal B} = -2.9$, the parity-odd components of the gravitational wave bispectrum are given by \cite{Shiraishi:2012sn}
\begin{eqnarray}
 B_{{\bf k}_1 {\bf k}_2 {\bf k}_3}^{\lambda_1 \lambda_2 \lambda_3} 
&\approx& 
0.6 \left[\frac{\ln(T_B / T_\nu )}{4\pi \rho_{\gamma,0}} \right]^3 
 k_*^{3}
  P_B (k_*)
\left[ 
P_B(k_1) \frac{\lambda_2}{2} P_{\cal B}(k_2)
+ 5~{\rm perms}
 \right]
 \nonumber \\
&&
\times e_{ij}^{(-\lambda_1)}(\hat{\bf k}_1) e_{jk}^{(-\lambda_2)}(\hat{\bf k}_2) e_{ki}^{(-\lambda_3)}(\hat{\bf k}_3)~,
\end{eqnarray}
where $k_* = 10~{\rm Mpc}^{-1}$ is a normalization scale. This has peaks at the squeezed limit due to the localized structure of gravitational waves \eqref{eq:GW_helical}; the resultant CMB bispectrum is thus also peaked on squeezed triangles, i.e., $\ell_1 \approx \ell_2 \gg \ell_3$. Following the squeezed-type normalization \eqref{eq:fnl_def_sq}, the nonlinearity parameter is defined by the combination of the magnetic field strengths and the energy scale of the PMF generation epoch, reading
\begin{eqnarray}
f_{\rm NL}^{H} \equiv 8 \left(\frac{B_{1}}{\rm 1~nG} \right)^4 
\left(\frac{{\cal B}_{1}}{\rm 1~nG} \right)^2 
\left(\frac{\ln(T_B / T_\nu )}{\ln(10^{17})} \right)^3 ~.
\end{eqnarray}
The expected $1\sigma$ error in the ideal noiseless full-sky experiment becomes $\delta f_{\rm NL}^H = 1.3 \times 10^3$.

Figure~\ref{fig:alphaR_betaR} shows, as expected, that the $\alpha^R$ spectrum of this model takes its largest contribution from the $n = 1$ squeezed mode. We already showed, when discussing bispectrum reconstruction, that squeezed 
modes are totally suppressed in the data, hence our final constraint is again fully consistent with Gaussianity, with $f_{\rm NL}$ well within the $1\sigma$ interval:
\begin{eqnarray}
f_{\rm NL}^H = (-0.6 \pm 1.5) \times 10^3 \ \ (68\% {\rm CL}) ~.
\end{eqnarray}
Also for this model, the central value in the raw map analysis is only very slightly different, namely $f_{\rm NL}^H = (-0.9 \pm 1.5) \times 10^3$. Assuming PMF generation at GUT energy scale, i.e., $T_B / T_\nu \approx 10^{17}$, and positiveness of $P_{\cal B}$, the $2\sigma$ constraint yields the upper bound: $B_1^{2/3} {\cal B}_1^{1/3} < 2.6$~nG. 

\section{Conclusion}

We have presented the first constraints on parity-odd NG by means of a modal bispectrum analysis of WMAP9 data. After validating our bispectrum estimation pipeline on simulations including realistic experimental features, we 
initially extracted a model-independent smoothed reconstruction of the data bispectrum in the parity-odd domain. Both are our modal expansion and bispectrum reconstruction shows consistency with Gaussianity, and larger fluctuations in the 
equilateral limit (although still consistent with $0$ within error bars).

Fitting modal expansion of theoretical bispectrum templates to our reconstructed data bispectrum, we have constrained three specific models giving rise to parity-odd NG, namely a Weyl cubic action, a rolling pseudoscalar and large-scale helical PMFs. 
The nonlinearity parameters $f_{\rm NL}$ have been found to be well consistent with $0$, with equilateral-type models (Weyl and rolling pseudoscalar) producing slightly larger values than the squeezed PMF template, as expected from the model-independent part of the analysis. 

In the future, {\it Planck} and possibly some proposed surveys like PRISM \cite{Andre:2013afa, Andre:2013nfa} will provide accurate CMB polarization data. As shown in refs.~\cite{Shiraishi:2013vha, Shiraishi:2013kxa}, including polarization
 in the estimators can generate a several-fold increase in the signal-to-noise ratios of parity-odd tensor bispectra. A joint analysis including temperature, E-mode and B-mode bispectra from the {\it Planck} dataset will provide interesting, more stringent constraints of these models
 in the near future.

\acknowledgments
We thank Paul Shellard for useful discussions. MS is supported in part by a Grant-in-Aid for JSPS Research under Grant No.~25-573. This work is supported in part by the ASI/INAF Agreement I/072/09/0 for the Planck LFI Activity of Phase E2. Some results in this paper have been obtained by use of the HEALPix package \cite{Gorski:2004by,HEALpix} and the dataset in the Lambda website \cite{Lambda}.


\bibliography{paper}

\providecommand{\href}[2]{#2}\begingroup\raggedright\begin{thebibliography}{10}

\bibitem{Bennett:2012zja}
{\bf WMAP} Collaboration, C.~Bennett {\em et.~al.}, {\it {Nine-Year Wilkinson
  Microwave Anisotropy Probe (WMAP) Observations: Final Maps and Results}},
  {\em Astrophys.J.Suppl.} {\bf 208} (2013) 20,
  [\href{http://xxx.lanl.gov/abs/1212.5225}{{\tt arXiv:1212.5225}}].

\bibitem{Ade:2013ydc}
{\bf Planck Collaboration} Collaboration, P.~Ade {\em et.~al.}, {\it {Planck
  2013 Results. XXIV. Constraints on primordial non-Gaussianity}},
  \href{http://xxx.lanl.gov/abs/1303.5084}{{\tt arXiv:1303.5084}}.

\bibitem{Babich:2004yc}
D.~Babich and M.~Zaldarriaga, {\it {Primordial bispectrum information from CMB
  polarization}},  {\em Phys.Rev.} {\bf D70} (2004) 083005,
  [\href{http://xxx.lanl.gov/abs/astro-ph/0408455}{{\tt astro-ph/0408455}}].

\bibitem{Yadav:2007rk}
A.~P. Yadav, E.~Komatsu, and B.~D. Wandelt, {\it {Fast Estimator of Primordial
  Non-Gaussianity from Temperature and Polarization Anisotropies in the Cosmic
  Microwave Background}},  {\em Astrophys.J.} {\bf 664} (2007) 680--686,
  [\href{http://xxx.lanl.gov/abs/astro-ph/0701921}{{\tt astro-ph/0701921}}].

\bibitem{Lue:1998mq}
A.~Lue, L.-M. Wang, and M.~Kamionkowski, {\it {Cosmological signature of new
  parity violating interactions}},  {\em Phys.Rev.Lett.} {\bf 83} (1999)
  1506--1509, [\href{http://xxx.lanl.gov/abs/astro-ph/9812088}{{\tt
  astro-ph/9812088}}].

\bibitem{Maldacena:2011nz}
J.~M. Maldacena and G.~L. Pimentel, {\it {On graviton non-Gaussianities during
  inflation}},  {\em JHEP} {\bf 1109} (2011) 045,
  [\href{http://xxx.lanl.gov/abs/1104.2846}{{\tt arXiv:1104.2846}}].

\bibitem{Soda:2011am}
J.~Soda, H.~Kodama, and M.~Nozawa, {\it {Parity Violation in Graviton
  Non-gaussianity}},  {\em JHEP} {\bf 1108} (2011) 067,
  [\href{http://xxx.lanl.gov/abs/1106.3228}{{\tt arXiv:1106.3228}}].

\bibitem{Shiraishi:2011st}
M.~Shiraishi, D.~Nitta, and S.~Yokoyama, {\it {Parity Violation of Gravitons in
  the CMB Bispectrum}},  {\em Prog.Theor.Phys.} {\bf 126} (2011) 937--959,
  [\href{http://xxx.lanl.gov/abs/1108.0175}{{\tt arXiv:1108.0175}}].

\bibitem{Shiraishi:2012sn}
M.~Shiraishi, {\it {Parity violation of primordial magnetic fields in the CMB
  bispectrum}},  {\em JCAP} {\bf 1206} (2012) 015,
  [\href{http://xxx.lanl.gov/abs/1202.2847}{{\tt arXiv:1202.2847}}].

\bibitem{Zhu:2013fja}
T.~Zhu, W.~Zhao, Y.~Huang, A.~Wang, and Q.~Wu, {\it {Effects of parity
  violation on non-gaussianity of primordial gravitational waves in
  Ho\v{r}ava-Lifshitz gravity}},  {\em Phys.Rev.} {\bf D88} (2013) 063508,
  [\href{http://xxx.lanl.gov/abs/1305.0600}{{\tt arXiv:1305.0600}}].

\bibitem{Cook:2013xea}
J.~L. Cook and L.~Sorbo, {\it {An inflationary model with small scalar and
  large tensor nongaussianities}},  {\em JCAP} {\bf 1311} (2013) 047,
  [\href{http://xxx.lanl.gov/abs/1307.7077}{{\tt arXiv:1307.7077}}].

\bibitem{Shiraishi:2013kxa}
M.~Shiraishi, A.~Ricciardone, and S.~Saga, {\it {Parity violation in the CMB
  bispectrum by a rolling pseudoscalar}},  {\em JCAP} {\bf 1311} (2013) 051,
  [\href{http://xxx.lanl.gov/abs/1308.6769}{{\tt arXiv:1308.6769}}].

\bibitem{Kamionkowski:2010rb}
M.~Kamionkowski and T.~Souradeep, {\it {The Odd-Parity CMB Bispectrum}},  {\em
  Phys.Rev.} {\bf D83} (2011) 027301,
  [\href{http://xxx.lanl.gov/abs/1010.4304}{{\tt arXiv:1010.4304}}].

\bibitem{Shiraishi:2014roa}
M.~Shiraishi, M.~Liguori, and J.~R. Fergusson, {\it {General parity-odd CMB
  bispectrum estimation}},  {\em JCAP} {\bf 1405} (2014) 008,
  [\href{http://xxx.lanl.gov/abs/1403.4222}{{\tt arXiv:1403.4222}}].

\bibitem{Fergusson:2009nv}
J.~Fergusson, M.~Liguori, and E.~Shellard, {\it {General CMB and Primordial
  Bispectrum Estimation I: Mode Expansion, Map-Making and Measures of $F_{\rm
  NL}$}},  {\em Phys.Rev.} {\bf D82} (2010) 023502,
  [\href{http://xxx.lanl.gov/abs/0912.5516}{{\tt arXiv:0912.5516}}].

\bibitem{Fergusson:2010dm}
J.~Fergusson, M.~Liguori, and E.~Shellard, {\it {The CMB Bispectrum}},  {\em
  JCAP} {\bf 1212} (2012) 032, [\href{http://xxx.lanl.gov/abs/1006.1642}{{\tt
  arXiv:1006.1642}}].

\bibitem{Liguori:2010hx}
M.~Liguori, E.~Sefusatti, J.~R. Fergusson, and E.~Shellard, {\it {Primordial
  non-Gaussianity and Bispectrum Measurements in the Cosmic Microwave
  Background and Large-Scale Structure}},  {\em Adv.Astron.} {\bf 2010} (2010)
  980523, [\href{http://xxx.lanl.gov/abs/1001.4707}{{\tt arXiv:1001.4707}}].

\bibitem{Fergusson:2011sa}
J.~R. Fergusson and E.~P.~S. Shellard, {\it {Optimal Polyspectra Estimation}},
  \href{http://xxx.lanl.gov/abs/1105.2791}{{\tt arXiv:1105.2791}}.

\bibitem{Fergusson:2014gea}
J.~Fergusson, {\it {Efficient optimal non-Gaussian CMB estimators with
  polarisation}},  \href{http://xxx.lanl.gov/abs/1403.7949}{{\tt
  arXiv:1403.7949}}.

\bibitem{Komatsu:2003iq}
E.~Komatsu, D.~N. Spergel, and B.~D. Wandelt, {\it {Measuring primordial
  non-Gaussianity in the cosmic microwave background}},  {\em Astrophys.J.}
  {\bf 634} (2005) 14--19,
  [\href{http://xxx.lanl.gov/abs/astro-ph/0305189}{{\tt astro-ph/0305189}}].

\bibitem{Komatsu:2003fd}
{\bf WMAP Collaboration} Collaboration, E.~Komatsu {\em et.~al.}, {\it {First
  year Wilkinson Microwave Anisotropy Probe (WMAP) observations: tests of
  gaussianity}},  {\em Astrophys.J.Suppl.} {\bf 148} (2003) 119--134,
  [\href{http://xxx.lanl.gov/abs/astro-ph/0302223}{{\tt astro-ph/0302223}}].

\bibitem{Creminelli:2005hu}
P.~Creminelli, A.~Nicolis, L.~Senatore, M.~Tegmark, and M.~Zaldarriaga, {\it
  {Limits on non-gaussianities from wmap data}},  {\em JCAP} {\bf 0605} (2006)
  004, [\href{http://xxx.lanl.gov/abs/astro-ph/0509029}{{\tt
  astro-ph/0509029}}].

\bibitem{Yadav:2007ny}
A.~P. Yadav, E.~Komatsu, B.~D. Wandelt, M.~Liguori, F.~K. Hansen, {\em
  et.~al.}, {\it {Fast Estimator of Primordial Non-Gaussianity from Temperature
  and Polarization Anisotropies in the Cosmic Microwave Background II: Partial
  Sky Coverage and Inhomogeneous Noise}},  {\em Astrophys.J.} {\bf 678} (2008)
  578--582, [\href{http://xxx.lanl.gov/abs/0711.4933}{{\tt arXiv:0711.4933}}].

\bibitem{Yadav:2007yy}
A.~P. Yadav and B.~D. Wandelt, {\it {Evidence of Primordial Non-Gaussianity
  (f(NL)) in the Wilkinson Microwave Anisotropy Probe 3-Year Data at
  2.8sigma}},  {\em Phys.Rev.Lett.} {\bf 100} (2008) 181301,
  [\href{http://xxx.lanl.gov/abs/0712.1148}{{\tt arXiv:0712.1148}}].

\bibitem{Komatsu:2008hk}
{\bf WMAP Collaboration} Collaboration, E.~Komatsu {\em et.~al.}, {\it
  {Five-Year Wilkinson Microwave Anisotropy Probe (WMAP) Observations:
  Cosmological Interpretation}},  {\em Astrophys.J.Suppl.} {\bf 180} (2009)
  330--376, [\href{http://xxx.lanl.gov/abs/0803.0547}{{\tt arXiv:0803.0547}}].

\bibitem{Senatore:2009gt}
L.~Senatore, K.~M. Smith, and M.~Zaldarriaga, {\it {Non-Gaussianities in Single
  Field Inflation and their Optimal Limits from the WMAP 5-year Data}},  {\em
  JCAP} {\bf 1001} (2010) 028, [\href{http://xxx.lanl.gov/abs/0905.3746}{{\tt
  arXiv:0905.3746}}].

\bibitem{Smith:2006ud}
K.~M. Smith and M.~Zaldarriaga, {\it {Algorithms for bispectra: Forecasting,
  optimal analysis, and simulation}},  {\em Mon.Not.Roy.Astron.Soc.} {\bf 417}
  (2011) 2--19, [\href{http://xxx.lanl.gov/abs/astro-ph/0612571}{{\tt
  astro-ph/0612571}}].

\bibitem{Liguori:2014}
M.~{Liguori}, M.~{Shiraishi}, J.~{Fergusson}, and E.~{Shellard} {\em In prep.}
  (2014).

\bibitem{Hinshaw:2012aka}
{\bf WMAP} Collaboration, G.~Hinshaw {\em et.~al.}, {\it {Nine-Year Wilkinson
  Microwave Anisotropy Probe (WMAP) Observations: Cosmological Parameter
  Results}},  {\em Astrophys.J.Suppl.} {\bf 208} (2013) 19,
  [\href{http://xxx.lanl.gov/abs/1212.5226}{{\tt arXiv:1212.5226}}].

\bibitem{Lambda}
\url{http://lambda.gsfc.nasa.gov}.

\bibitem{Komatsu:2010fb}
{\bf WMAP Collaboration} Collaboration, E.~Komatsu {\em et.~al.}, {\it
  {Seven-Year Wilkinson Microwave Anisotropy Probe (WMAP) Observations:
  Cosmological Interpretation}},  {\em Astrophys.J.Suppl.} {\bf 192} (2011) 18,
  [\href{http://xxx.lanl.gov/abs/1001.4538}{{\tt arXiv:1001.4538}}].

\bibitem{Shiraishi:2010kd}
M.~Shiraishi, D.~Nitta, S.~Yokoyama, K.~Ichiki, and K.~Takahashi, {\it {CMB
  Bispectrum from Primordial Scalar, Vector and Tensor non-Gaussianities}},
  {\em Prog.Theor.Phys.} {\bf 125} (2011) 795--813,
  [\href{http://xxx.lanl.gov/abs/1012.1079}{{\tt arXiv:1012.1079}}].

\bibitem{Shiraishi:2010sm}
M.~Shiraishi, S.~Yokoyama, D.~Nitta, K.~Ichiki, and K.~Takahashi, {\it
  {Analytic formulae of the CMB bispectra generated from non-Gaussianity in the
  tensor and vector perturbations}},  {\em Phys.Rev.} {\bf D82} (2010) 103505,
  [\href{http://xxx.lanl.gov/abs/1003.2096}{{\tt arXiv:1003.2096}}].

\bibitem{Barnaby:2012xt}
N.~Barnaby, J.~Moxon, R.~Namba, M.~Peloso, G.~Shiu, {\em et.~al.}, {\it
  {Gravity waves and non-Gaussian features from particle production in a sector
  gravitationally coupled to the inflaton}},  {\em Phys.Rev.} {\bf D86} (2012)
  103508, [\href{http://xxx.lanl.gov/abs/1206.6117}{{\tt arXiv:1206.6117}}].

\bibitem{Caprini:2003vc}
C.~Caprini, R.~Durrer, and T.~Kahniashvili, {\it {The Cosmic microwave
  background and helical magnetic fields: The Tensor mode}},  {\em Phys.Rev.}
  {\bf D69} (2004) 063006,
  [\href{http://xxx.lanl.gov/abs/astro-ph/0304556}{{\tt astro-ph/0304556}}].

\bibitem{Andre:2013afa}
{\bf PRISM Collaboration} Collaboration, P.~Andre {\em et.~al.}, {\it {PRISM
  (Polarized Radiation Imaging and Spectroscopy Mission): A White Paper on the
  Ultimate Polarimetric Spectro-Imaging of the Microwave and Far-Infrared
  Sky}},  \href{http://xxx.lanl.gov/abs/1306.2259}{{\tt arXiv:1306.2259}}.

\bibitem{Andre:2013nfa}
{\bf PRISM Collaboration} Collaboration, P.~Andre {\em et.~al.}, {\it {PRISM
  (Polarized Radiation Imaging and Spectroscopy Mission): An Extended White
  Paper}},  {\em JCAP} {\bf 1402} (2014) 006,
  [\href{http://xxx.lanl.gov/abs/1310.1554}{{\tt arXiv:1310.1554}}].

\bibitem{Shiraishi:2013vha}
M.~Shiraishi, {\it {Polarization bispectrum for measuring primordial magnetic
  fields}},  {\em JCAP} {\bf 1311} (2013) 006,
  [\href{http://xxx.lanl.gov/abs/1308.2531}{{\tt arXiv:1308.2531}}].

\bibitem{Gorski:2004by}
K.~Gorski, E.~Hivon, A.~Banday, B.~Wandelt, F.~Hansen, {\em et.~al.}, {\it
  {HEALPix - A Framework for high resolution discretization, and fast analysis
  of data distributed on the sphere}},  {\em Astrophys.J.} {\bf 622} (2005)
  759--771, [\href{http://xxx.lanl.gov/abs/astro-ph/0409513}{{\tt
  astro-ph/0409513}}].

\bibitem{HEALpix}
\url{http://healpix.jpl.nasa.gov/}.

\end{thebibliography}\endgroup
\end{document}